\documentclass[aps,prl,showpacs,twocolumn,amsmath,amssymb]{revtex4}
\usepackage{graphicx}
\usepackage{amsmath}
\usepackage{amsfonts}

\begin{document}

\title{Guided Quasicontinuous Atom Laser}

\date{\today}

\author{W. Guerin}
\email[Email address: ]{William.Guerin@institutoptique.fr}
\homepage[Electronic address: ]{www.atomoptic.fr}
\author{J.-F. Riou}
\author{J. P. Gaebler}
\altaffiliation[Present address: ]{JILA, University of Colorado,
440 UCB, Boulder, CO 80309-0440, U.S.A.}
\author{V. Josse}
\author{P. Bouyer}
\author{A. Aspect}

\affiliation{Laboratoire Charles Fabry de l'Institut d'Optique, CNRS et Universit\'{e} Paris Sud 11\\
Campus Polytechnique, RD 128, 91127 Palaiseau, France}

\begin{abstract}
We report the first realization of a guided quasicontinuous atom
laser by rf outcoupling a BEC, from a hybrid optomagnetic trap
into a horizontal atomic waveguide. This configuration allows us
to cancel the acceleration due to gravity and keep the de~Broglie
wavelength constant at 0.5~$\mu$m during 0.1~s of propagation. We
also show that our configuration, equivalent to \emph{pigtailing}
an optical fiber to a (photon) semiconductor laser, ensures an
intrinsically good transverse mode matching.
\end{abstract}


\maketitle

The Bose-Einstein condensation of atoms in the lowest level of a
trap represents the matter-wave analog to the accumulation of
photons in a single mode of a laser cavity. In analogy to photonic
lasers, atom lasers can be obtained by outcoupling from a trapped
Bose-Einstein condensate (BEC) to free space
\cite{laser_gallery,Bloch:1999,LeCoq:2001}. However, since atoms are
massive particles, gravity plays an important role in the laser
properties: in the case of rf outcouplers, it lies at the very heart
of the extraction process \cite{Gerbier:2001} and in general, the
beam is strongly accelerated downwards, causing a rapid decrease of
the de~Broglie wavelength. With the growing interest in coherent
atom sources for atom interferometry
\cite{bouyer:1997,Gupta:2002,Wang:2005} and new studies of quantum
transport phenomena
\cite{carusotto:2001,Leboeuf:2001,pavloff:2002,Cataliotti:2003,Paul:2005_1,Paul:2005_2,desordre:2005}
where large and well defined de Broglie wavelength are desirable, a
better control of the atomic motion during its propagation is
needed. One solution is to couple the atom laser into a horizontal
waveguide, so that the effect of gravity is canceled, leading to the
realization of a coherent matter wave with constant wavelength.

\begin{figure}[b]
    \centering
    \includegraphics{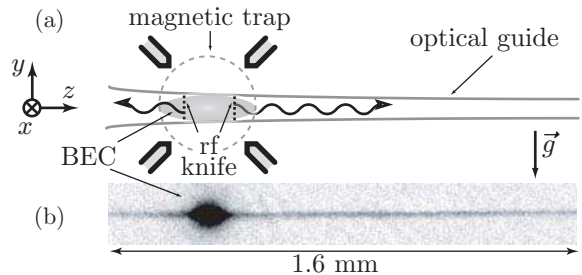}
    \caption{(a) Setup.
    The BEC is produced at the intersection of a magnetic trap and a
    horizontal elongated optical trap acting as a waveguide for
the atom laser. A ``rf knife" provides outcoupling into the
waveguide and an atom laser is
    emitted on both sides.
    (b) Absorption image (along $x$) of a guided atom laser after 100~ms
of outcoupling.} \label{fig_manip}
\end{figure}

We report in this letter on the realization of such a guided
quasicontinuous atom laser, where the coherent source, \textit{i.e.}
the trapped BEC, and the guide are {\em merged} together in a hybrid
combination of a magnetic Ioffe-Pritchard trap and a horizontally
elongated far off-resonance optical trap constituting an atomic
waveguide (see Fig. \ref{fig_manip}). The BEC, in a state sensitive
to both trapping potentials, is submitted to a rf outcoupler
yielding atoms in a state sensitive only to the optical potential,
resulting in an atom laser propagating along the weak confining axis
of the optical trap. In addition to canceling the effect of gravity,
this configuration has several advantages. Firstly, coupling into a
guide from a BEC rather than from a thermal sample
\cite{guide_thermique} allows us to couple a significant flux into a
small number of transverse modes of the guide. Secondly, the weak
longitudinal trapping potential of the guide can be compensated by
the antitrapping potential due to the second order Zeeman effect
acting onto the outcoupled atoms, resulting in an atom laser with a
quasiconstant de~Broglie wavelength. Thirdly, using an rf outcoupler
rather than releasing a BEC into a guide
\cite{guide_bec,desordre:2005} results into quasicontinuous
operation, thus insuring sharp linewidth, and gives a better control
on the beam parameters. Indeed, changing the frequency of the
outcoupler allows one to \emph{tune} the value of the de~Broglie
wavelength of the atom laser, and adjusting the rf coupler power
allows one to independently vary the atom-laser density from the
interacting regime to the noninteracting one \cite{footnote_Lahaye}.
In particular, those advantages opens new prospects for studying
quantum transport phenomena, as, for instance, quantum reflection
\cite{Shimizu:2001}, where interactions dramatically suppress the
reflection probability \cite{Pasquini:2006}. Finally, in spite of
the lensing effect due to the interaction of the atom laser with the
trapped BEC \cite{LeCoq:2001,Riou:2006}, adiabatic transverse mode
matching results into the excitation of only a small number of
transverse modes, and we discuss the possibility of achieving single
transverse mode operation.

Our setup \cite{fauquembergue:2005} produces magnetically trapped
cold clouds of $\mathrm{{}^{87}Rb}$ in the
$\left|F,m_{F}\right\rangle\!=\!\left|1,-1\right\rangle$ state.
After evaporative cooling to $1~\mu$K, an optical guide produced by
120~mW of Nd:YAG laser ($\lambda\!=\!1064$~nm) focussed on a waist
of 30~$\mu$m is superimposed along the $z$ direction and after a
final evaporation ramp of 6~s \cite{footnote_evap}, Bose-Einstein
condensation is directly obtained in the optomagnetic trap. We
estimate the condensed fraction to 80\%
($T\approx\!0.4T_\mathrm{c}\!\approx\!150$~nK) with 10$^5$ atoms in
the BEC. In this hybrid trap, the optical guide ensures a tight
transverse confinement, with oscillation frequencies
$\omega_{x,y}/2\pi\!=\!\omega_{\perp}/2\pi\!=\!360$~Hz, large
compared to those of the magnetic trap
($\omega_x^\mathrm{m}/2\pi\!=\!8$~Hz and
$\omega_y^m/2\pi\!=\!35$~Hz). In contrast, the confinement along the
$z$ axis is due to the shallow magnetic trap with an oscillation
frequency $\omega_z^m/2\pi\!=\!35$~Hz. The chemical potential is
then $\mu_\mathrm{BEC}/h\simeq 3.2$~kHz and the Thomas-Fermi radii
are $R_z\!=\!25~\mu$m and $R_\bot\!=\!2.4~\mu$m. The guided atom
laser is obtained by rf-induced magnetic transition
\cite{Bloch:1999} between the $\left|1,-1\right\rangle$ state and
the $\left|1,0\right\rangle$ state, which is submitted to the same
transverse confinement due to the optical guide, but is not
sensitive (at first order) to the magnetic trapping. We thus obtain
a guided coherent matter wave propagating along the optical guide
[Fig. \ref{fig_manip}(b)]. This configuration, where the optical
guide dominates the transverse trapping of both the source BEC and
the atom laser, enables to collect the outcoupled atoms into the
guide with 100$\%$ efficiency.

\begin{figure}[t]
    \centering
    \includegraphics{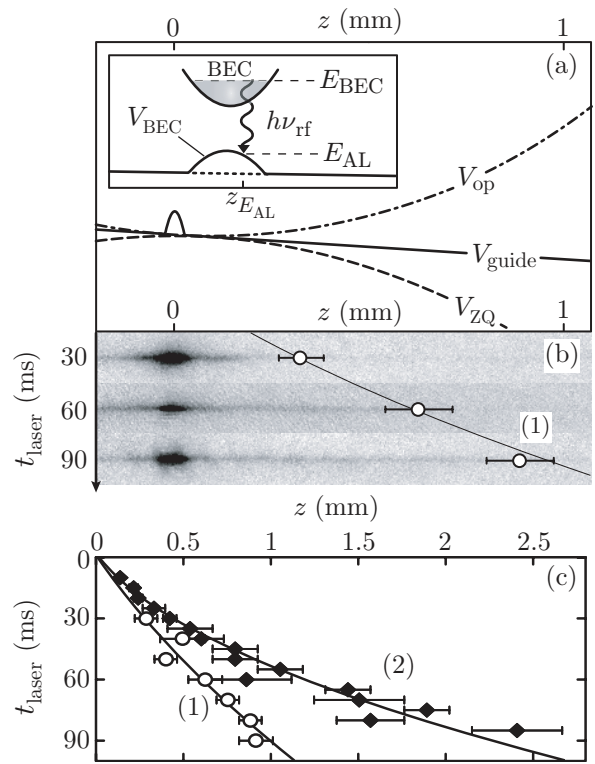}
    \caption{Longitudinal dynamics of the guided atom laser. (a) Longitudinal potential $V_{\mathrm{guide}}+V_{\mathrm{BEC}}$, sum of the quadratic Zeeman (dashed), optical (dash-dot) and BEC mean-field (inset) potentials.
    (b) Guided atom laser after different lasing times $t_{\mathrm{laser}}$. These images allow us to determine the wavefront position (estimated error bars are shown).
    (c) Wavefront position versus $t_\mathrm{laser}$ for two different adjustments of the optical potential. Each set of data is fitted by a second degree polynomial, yielding the same initial velocity  $v_0=9\,\pm\,2$~mm.s$^{-1}$, and different accelerations  $a_1=0.07\,\pm\,0.06$~m.s$^{-2}$ (1) and  $a_2=0.36\,\pm\,0.04$~m.s$^{-2}$
    (2).}
    \label{fig_propag}
\end{figure}

As explained below, the propagation of the guided atom laser,
after leaving the region of interaction with the remaining BEC, is
dominated by a potential $V_{\mathrm{guide}}(z)$ resulting from
the repulsive second order Zeeman effect
$V_{\mathrm{ZQ}}(z)\!=\!-m\omega_{\mathrm{ZQ}}^2(z-z_{\mathrm{m}})^2/2$
and the weakly trapping optical potential
$V_{\mathrm{op}}(z)\!=\!m\omega_{\mathrm{op}}^2(z-z_0)^2/2$, where
$z_{\mathrm{m}}$ and $z_0$ are respectively the magnetic and
optical traps centers relative to the BEC center
\cite{footnote_sag}. For our parameters the curvatures of
$V_{\mathrm{ZQ}}(z)$ and $V_{\mathrm{op}}(z)$ cancel each other
($\omega_{\mathrm{op}}/2\pi\!\simeq\!\omega_{\mathrm{ZQ}}/2\pi
\!=\!2$~Hz), so that $V_\mathrm{guide}(z)$ is nearly linear, with
a slope corresponding to an acceleration
$a_\mathrm{guide}\!=\!\omega_{\mathrm{op}}^{2} z_{0}$, several
orders of magnitude smaller than gravity [Fig.~\ref{fig_propag}].
Then the atom-laser velocity remains almost constant at
$v=9$~mm.s$^{-1}$, corresponding to a de~Broglie wavelength
$\lambda_{\mathrm{db}}=h/mv$ of $0.5~\mu$m.

Besides its de Broglie wavelength, an atom laser is characterized
by its flux. In quasicontinuous rf outcoupling and in the weak
coupling regime \cite{Gerbier:2001,Robins:2005}, this flux can be
controlled by adjusting the rf power. We work at a flux
$\mathcal{F}=5\times10^5$~at.s$^{-1}$ which is appropriate for
efficient absorption imaging of the atom laser. The dimensionless
parameter $n_{\mathrm{1D}}a_\mathrm{s}$ characterizing the
interactions \cite{Jackson:1998} is about $0.25$. In this
expression, $a_\mathrm{s}=5.3$~nm is the (3D) atomic scattering
length and $n_{\mathrm{1D}}$ is the linear density
($n_{\mathrm{1D}}= \mathcal{F}/v \simeq 45$~at.$\mu$m$^{-1}$ at
$v=9$~mm.s$^{-1}$). For $n_{\mathrm{1D}}a_\mathrm{s}\!<\!1$ we are
in the ``1D mean field" regime \cite{Menotti:2002}, where the
mean-field intralaser interaction may influence the longitudinal
dynamics but not the transverse one.

Our modeling of the dynamics of the guided atom laser is based on
the formalism used in \cite{Jackson:1998}. The strong transverse
confinement allows us to assume that the quantized transverse
dynamics adiabatically follows the slowly varying transverse
potential as the laser propagates along the $z$ axis. In this
``quasi-1D regime", the laser wave function takes the form:
\begin{equation}\Psi (\vec{r},t)=\phi(z,t)
\psi_{\perp}(\vec{r}_{\perp},z)
\label{1Dwave}
\end{equation}
with the normalization $\int |\psi_{\perp}|^2d\vec{r}_{\perp}=1$
so that the linear density is $n_{\mathrm{1D}}\!=\!\int
|\Psi|^2d\vec{r}_{\perp}\!=\!|\phi(z,t)|^2$. In the following we
will assume that $\psi_{\perp}(\vec{r}_{\perp},z)$ is the ground
state of the local transverse potential including the mean-field
interaction due to the BEC, so that it matches perfectly the BEC
transverse shape in the overlap region and evolves smoothly to a
gaussian afterwards. The longitudinal dynamics can then be
described in terms of hydrodynamical equations, bearing on
$n_{\mathrm{1D}}$ and the phase velocity
$v\!=\!\hbar\nabla\mathcal{S}/m$ such that
$\phi\!=\!\sqrt{n_{\mathrm{1D}}}~e^{i\mathcal{S}}$. In the
stationary regime, for an atom laser of energy $E_{\mathrm{AL}}$,
these equations reduce to the atomic flux and energy
conservations:
\begin{gather}
n_{\mathrm{1D}}(z)\: v(z) = \mathcal{F}\:,\label{flux}\\
\frac{1}{2}mv(z)^2+V_{\mathrm{guide}}(z)+\mu(z) =
E_{\mathrm{AL}}\:.\label{Bernouilli}
\end{gather}
The quantity $\mu(z)$ is an effective local chemical potential which
takes into account both intralaser interaction and transverse
confinement \cite{Jackson:1998}. Inside the BEC, $\mu(z)$ is
dominated by the interaction with the trapped BEC and we can rewrite
$\mu(z)\!=\!V_{\mathrm{BEC}}(z)\!=\!\mu_{\mathrm{BEC}}(1-z^2/R_z^2)$.
Outside the BEC and in the ``1D mean field" regime, one has
$\mu(z)\!=\!\hbar\omega_\perp(1+2a_\mathrm{s} n_{\mathrm{1D}}(z))$.

To write Eq. (\ref{Bernouilli}), we have neglected the
longitudinal quantum pressure since the density $n_{\mathrm{1D}}$
varies smoothly along $z$. With this simplification, Eqs.
(\ref{flux}) and (\ref{Bernouilli}) are equivalent  to the
standard WKB approximation. The amplitude of $\phi(z,t)$ is
determined by the flux $\mathcal{F}$ [Eq. (\ref{flux})] and its
phase $\mathcal{S}(z)$ can be derived from the classical motion of
an atom of energy $E_{\mathrm{AL}}$ submitted to the 1D potential
$V_{\mathrm{AL}}(z)\!=\!V_{\mathrm{guide}}(z)+\mu(z)$. The
parameters $E_{\mathrm{AL}}$ and $\mathcal{F}$, determining the
atom-laser wave function, are fixed by the frequency and power of
the output coupler.

In the weak coupling regime, the coupling between the trapped BEC
and the continuum of propagating atom-laser wave functions can be
described by the Fermi Golden Rule (see \cite{Gerbier:2001} and
references \textit{therein}). The atom-laser energy is thus given
by the resonance condition
\begin{equation}
E_{\mathrm{AL}}=E_{\mathrm{BEC}}-h\nu_{\mathrm{rf}},
\label{energie}
\end{equation}
where $E_{\mathrm{BEC}}$ is the BEC energy, and the coupling rate,
which determines $\mathcal{F}$, depends on the overlap integral
between the BEC and the atom-laser wave functions. For a uniformly
accelerated atom laser, the longitudinal wave function $\phi(z,t)$
is an Airy function with a narrow lobe around the classical turning
point $z_{E_{\mathrm{AL}}}$, defined by $v(z_{E_\mathrm{AL}})\!=\!0$
in Eq. (\ref{Bernouilli}), and the overlap integral is proportional
to the BEC wave function at $z_{E_{\mathrm{AL}}}$
\cite{Gerbier:2001}. This can be interpreted by the so-called
Franck-Condon principle, which states that the rf coupler selects,
via the resonance condition, the atom laser extraction position
$z_{E_{\mathrm{AL}}}$ \cite{Band:1999}. In contrast to the case
where the atom laser is extracted by gravity, here the acceleration
due to $V_{\mathrm{guide}}(z)$ is small enough that the potential
$V_{\mathrm{AL}}(z)$ is dominated by the bump $V_{\mathrm{BEC}}(z)$
[Fig. \ref{fig_propag}(a)], so that there are two outcoupling points
corresponding to two atom lasers emitted on both sides of the
trapped condensate [Fig. \ref{fig_propag}(b)]. If the slope of the
potential $ma(z_{E_{\mathrm{AL}}})$ varies slowly around the
outcoupling point at the scale of the first lobe of the
corresponding Airy function, the atom-laser wave function can be
locally approximated by the Airy function and we can use the result
of \cite{Gerbier:2001} where gravity acceleration is replaced by
$a(z_{E_\mathrm{AL}})$:
\begin{equation} \label{franckcondon}
\mathcal{F}=
\frac{\pi\hbar\Omega^2_{\mathrm{rf}}}{2}\frac{n_{\mathrm{1D}}^\mathrm{{BEC}}(z_{E_{\mathrm{AL}}})}{ma(z_{E_{\mathrm{AL}}})}.
\end{equation}
Here $\Omega_{\mathrm{rf}}$ is the Rabi frequency characterizing
the rf coupling between the different atomic internal states, and
$n_{\mathrm{1D}}^{\mathrm{BEC}}(z)\!=\!\int
d\vec{r}_{\perp}\,\left|\psi_{\mathrm{BEC}}(\vec{r}_{\perp},z)\right|^2$
is the condensate linear density. More rigourously, one can solve
the Schr\"{o}dinger equation in a parabolic antitrapping potential
\cite{Fertig:1987}. We checked that the two calculations give the
same result when the local slope approximation is valid, and the
second approach is necessary only when the coupling is close to
the maximum of the potential bump. As expected, the flux is then
predicted to reach its maximum value.

The modeling above allows us to analyze our experimental data.
Firstly, for a Rabi frequency of
$\Omega_{\mathrm{rf}}/2\pi\!=\!40$~Hz, a BEC of
$N_{\mathrm{BEC}}\!\simeq\!10^5$~atoms and assuming a coupling at
about $5~\mu$m from the center of the BEC, Eq. (\ref{franckcondon})
gives $\mathcal{F}\!=\!5\times10^5$~at.s$^{-1}$, in agreement with
the observed decay of the atom number in the BEC. Secondly, this
modeling shows that with our parameters, the axial dynamics of the
atom laser associated to Eqs. (\ref{flux}) and (\ref{Bernouilli}) is
revealed by the propagation of the wavefront of the atom laser [Fig.
\ref{fig_propag}(b)]. Indeed, out of the region of overlap with the
trapped BEC, and for a coupling close to the potential maximum, the
atoms have a kinetic energy of the order of the BEC chemical
potential ($\mu_\mathrm{BEC}/h\!\simeq\!3.2$~kHz), large compared to
$\mu(z)$ ($\mu(z)/h\!\sim\!\omega_{\perp}/2\pi\!=\!360$~Hz). We can
thus neglect $\mu(z)$ in Eq. (\ref{Bernouilli}), and out of the BEC
the wavefront acceleration is dominated by $V_\mathrm{guide}(z)$,
while the atomic velocity just leaving the BEC is determined by
$V_\mathrm{BEC}(z_{E_\mathrm{AL}})$. For an outcoupling at the
center of the BEC, the expected value is
$v_0\!\simeq\!5.4$~mm.s$^{-1}$, somewhat less than the observed
value $v_0\!=\!9\!\pm\,2$~mm.s$^{-1}$. The discrepancy will be
discussed below.

We now turn to the transverse mode of the guided atom laser. To
characterize it, we measure the transverse energy using a
time-of-flight: after $60$~ms of propagation, the optical guide is
suddenly switched off and we measure the expansion along the $y$
axis. The evolution of the rms size is directly related to the
transverse kinetic energy according to
$\sigma(t)^2\!=\!\sigma_{0}^{2}+\!<\!v_{y}^{2}\!>\! t^{2}$, where
$\sigma_{0}$ is the resolution of the imaging system ($7.5~\mu$m)
which dominates the initial transverse size ($0.6~\mu$m). A fit
gives $<v_{y}^{2}> =\!4.5\!\pm\!0.2$~mm$^{2}$/s$^{2}$. Assuming
cylindrical symmetry, this corresponds to a total transverse energy
$E_\perp\!=\!(5.5\!\pm\!0.8) \hbar \omega_\perp$, \textit{i.e.} an
average excitation quantum number of 2 along each transverse
direction. This shows that only a few transverse modes are excited,
and we may wonder whether single transverse mode operation is
achievable.

Actually, we expect the atom laser to be outcoupled in its lowest
transverse mode. Indeed, the transverse potential experienced by an
atom in the atom laser has the same shape as the one experienced by
an atom of the BEC, \textit{i.e.}, in the Thomas-Fermi
approximation, quadratic trapping edges and a flat bottom of width
$2 R_\perp(z)$. As $z$ increases, this width decreases monotonically
to 0 until the point where the atom laser leaves the BEC and
experiences a pure harmonic potential. A numerical simulation shows
that this evolution is smooth enough to enable the transverse
atom-laser wave function $\psi_\perp(\vec{r_\perp},z)$ to
adiabatically adjust to the local ground state, resulting in the
prediction of almost single-mode emission. The observed multimode
behavior may be attributed to different experimental imperfections,
which can be fixed in future experiments. Firstly, if the magnetic
trap is not exactly centered on the optical guide, transverse mode
matching between the BEC and the guide is not perfect. Secondly,
excitation of higher transverse modes can be provoked by the
position noise of the guide (we observe a heating rate of 100~nK/s).
Finally, a numerical resolution of the coupled Gross-Pitaevskii
equations suggests that at our value of the atomic flux, the BEC
decay is not adiabatic enough \cite{Gerbier:2001} so that the
outcoupling could induce excitations inside the BEC and thus
increase the energy transferred to the atom laser. This might also
explain why the observed values of atom-laser velocity correspond to
an energy somewhat larger than $\mu_\mathrm{BEC}$.

In conclusion, we have demonstrated a scheme for efficiently
coupling a BEC into a waveguide. We have obtained a guided atom
laser with an almost constant de Broglie wavelength, at a value of
$0.5\:\mu$m, and by coupling near the boundary of the BEC it should
be possible to obtain even larger de Broglie wavelengths. Such
values are of interest for experiments in atom interferometry as,
for instance, the coherent splitting at the crossing of two
matterwave guides \cite{Houde:2000,Dumke:2002}, which could be
implemented in miniaturized components \cite{Birkl:2001}.
Furthermore, as the atomic wavelength reaches values similar to
visible light wavelength, transport properties through wells,
barriers or disordered structures engineered with light should enter
the quantum regime
\cite{carusotto:2001,Leboeuf:2001,pavloff:2002,Paul:2005_1,Cataliotti:2003,Paul:2005_2,desordre:2005}.
Also the control of the atom-laser flux offers the possibility to
tune the amount of interaction inside the guided atom-laser beam.
For instance, the possibility of combining a large and well defined
de Broglie wavelength together with a density small enough to
suppress interactions, should provide the conditions to observe
Anderson-like localization \cite{Paul:2005_2}. On the other hand,
the interacting regime should allow investigation of effects such as
the breakdown of superfluidity through obstacles
\cite{pavloff:2002,Cataliotti:2003}, or nonlinear resonant transport
\cite{carusotto:2001,Paul:2005_1}. We thus believe that our scheme
constitutes a very promising tool for further development of
coherent guided atom optics.

\begin{acknowledgments}
The authors would like to thank M. Fauquembergue and Y. Le Coq for
their help at the early stages of the experiment and D. Cl\'{e}ment
for fruitful discussions. The Groupe d'Optique Atomique is a member
of IFRAF. This work is supported by CNES (DA:10030054), DGA
(contracts 9934050 and 0434042), LNE, EU (grants IST-2001-38863,
MRTN-CT-2003-505032 and FINAQS STREP) and ESF (BEC2000+ and
QUDEDIS).
\end{acknowledgments}


\begin{thebibliography}{38}
\bibitem{footnote_evap} Because of the strong rf, evaporated atoms are in the antitrapping $|1,1\rangle$ state and escape rapidly along the guide axis.
\bibitem{footnote_sag} Due to the optical trap, the BEC is shifted
from the magnetic trap center by
$z_{\mathrm{m}}=-(\omega_{\mathrm{op}}/\omega_z^m)^2z_0$.
\end{thebibliography}
\end{document}